\documentclass[12pt]{emulateapj}

\shorttitle{Collisionally excited filaments in HH~1/2}
\shortauthors{Raga et al.}

\begin{document}

\title{Collisionally excited filaments in HST H$\alpha$ and H$\beta$ images \\
of HH~1/2}

\author{A. C. Raga\altaffilmark{1}, B. Reipurth\altaffilmark{2},
A. Castellanos-Ram\'\i rez\altaffilmark{1}, Hsin-Fang Chiang\altaffilmark{2}, J. Bally\altaffilmark{3}}
\altaffiltext{1}{Instituto de Ciencias Nucleares, Universidad
Nacional Aut\'onoma de M\'exico, Ap. 70-543, 04510 D.F., M\'exico}
\altaffiltext{2}{Institute for Astronomy, University of Hawaii at Manoa, Hilo, HI 96720, USA}
\altaffiltext{3}{Center for Astrophysics and Space Astronomy, University of Colorado, UCB 389,
Boulder, CO 80309, USA}

\email{raga@nucleares.unam.mx}

\begin{abstract}
We present new H$\alpha$ and H$\beta$ images of the HH~1/2 system, and we find that the
H$\alpha$/H$\beta$ ratio has high values in ridges along the leading edges of the HH~1
bow shock and of the brighter condensations of HH~2. These ridges have H$\alpha$/H$\beta=4\to 6$,
which is consistent with collisional excitation from the $n=1$ to the $n=3$ and 4 levels
of hydrogen in a gas of temperatures $T=1.5\to 10\times 10^4$~K. This is therefore the first
direct proof that the collisional excitation/ionization region of hydrogen right behind Herbig-Haro
shock fronts is detected.
\end{abstract}


\section{Introduction}

HH~1 and 2 were the first Herbig-Haro (HH) objects to be discovered
(Herbig 1951; Haro 1952), and have played an important role in the field of
HH objects (see the historical review of Raga et al. 2011). For example, HST images (Schwartz et al. 1993;
Hester et al. 1998), proper motions (ground based: Herbig \& Jones 1981; HST: Bally et al. 2002; IR:
Noriega-Crespo et al. 1997; radio: Rodr\'\i guez et al. 2000), and detections
in radio continuum (Pravdo et al. 1985), UV (Ortolani \& D'Odorico 1980) and X-rays (Pravdo et al. 2001)
were first obtained for HH~1 and 2.

The HH~1/2 system has a central source detected in radio continuum (see, e.g.,
Rodr\'\i guez et al. 2000) and a bipolar jet system, with a NW jet (directed towards
HH~1) which is visible optically, and a SE jet (directed towards HH~2) visible only
in the IR (see Noriega-Crespo \& Raga 2012). HH~1 has a ``single bow shock'' morphology,
and HH~2 is a collection of condensations, some of them also with bow-shaped
morphologies (see, e.g., Bally et al. 2002). The emission-line structure of these
objects has been studied spectroscopically, with 1D (Solf, B\"ohm \& Raga
1988) and 2D (Solf et al. 1991; B\"ohm \& Solf 1992) coverage of the objects. It should be
pointed out that the HH~1/2 outflow lies very close to the plane of the sky, so that projection
effects do not have to be considered when interpreting the observations of these objects.

The spatial structure of the optical line emission has been studied at higher angular
resolution with HST images. Schwartz et al. (1993) obtained H$\alpha$, [S~II] 6716+6730
and [O~I] 6300 images. Later images of HH~1 and 2 were
all taken with filters isolating the H$\alpha$ and the red [S~II] lines (Bally et al. 2002;
Hartigan et al. 2011).

In the present paper we describe a pair of new HST images of HH~1 and 2 obtained
with filters isolating the H$\alpha$ and H$\beta$ lines. These images were obtained
in consecutive exposures, so that they are not affected by proper motions (which become
evident in HST observations of the HH~1/2 complex separated by more than a few weeks) nor by differences in
the pointing, and they therefore yield an accurate depiction of the spatial distribution
of the H$\alpha$/H$\beta$ ratio in these objects. These images show effects that
have not been detected before in ground based studies of the emission line structure
of HH~1 and 2 (see, e.g., Solf et al. 1991 and B\"ohm \& Solf 1992) nor in HST images
of other HH objects (since HST H$\beta$ images of HH objects have not been previously obtained).

The paper is organized as follows. The new HST images are described in section 2. The
spatial distribution of the H$\alpha$/H$\beta$ ratio, the line ratios as a function of H$\beta$ intensity
and the distribution functions of the line ratios are presented in section 3. Finally, an interpretation
of the results is presented in section 4.

\section{The observations}

The region around HH~1 and 2 was observed with the H$\alpha$ (F656N) and
H$\beta$ (F487N) filters on August 16, 2014 with the WFC3 camera on the HST. 
The H$\alpha$ image was obtained with a 2686 s exposure and the H$\beta$ image
with a slightly longer, 2798 s exposure. The images were reduced with the standard
pipeline, and a simple recognition/replacement algorithm was used to remove the
cosmic rays. The final images have $4130\times 4446$ pixels, with a pixel size of $0''.03962$.

The images contain only two stars: the Cohen-Schwartz star (Cohen \& Schwartz 1979)
and ``star no. 4'' of Strom et al. (1985).
These two stars have been used to determine astrometric positions in CCD images of the HH~1/2
region since the work of Raga et al. (1990), yielding better positions for HH~1 (which is closer
to the two stars) than for HH~2. We have carried out paraboloidal fits to the PSFs of these two stars,
and find no evidence for offsets and/or rotation, which shows the excellent tracking of the HST
during the single pointing in which the two images were obtained. Also, we have analyzed the
H$\alpha-$H$\beta$ difference images of the two stars, and find no offsets between the two frames.

The full H$\alpha$ frame is shown in Figure 1, as well as blow-ups of regions around HH~1 and HH~2 in
both H$\alpha$ and H$\beta$.
As seen in the top frame, the H$\alpha$ map shows the extended collection of HH~2 knots (to the SE)
and the more compact distribution of the HH~1 knots (towards the NW). The central frames of Figure 1 show
the H$\alpha$ emission of HH~2 (left) and HH~1 (right) at a larger scale. In the fainter H$\beta$ emission
(bottom frames of Figure 1) only the brighter regions of HH~1 and 2 are detected. We have defined two boxes
(labeled A and B in the bottom frame of Figure 1) enclosing the regions of the two objects which are
detected in H$\beta$. In the following section, the regions within these two boxes are used in order to study
the spatial dependence of the H$\alpha$/H$\beta$ ratio.

\begin{figure}
\centering
\includegraphics[width=8cm]{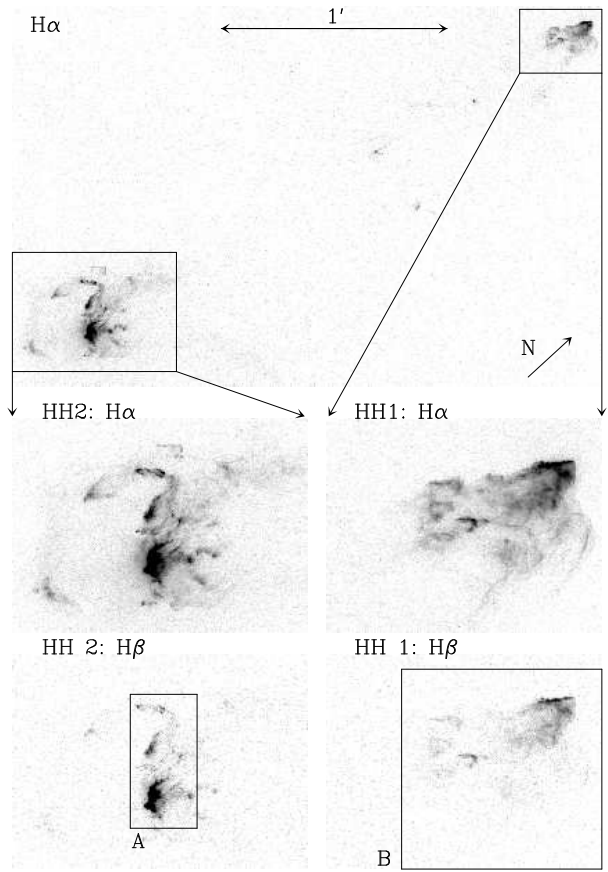}
\caption{Top: H$\alpha$ frame of HH~1 and 2 obtained with the WFC3
camera of the HST (the scale and orientation of the
images is shown). The central and the bottom frames show the H$\alpha$ and H$\beta$
images (respectively) of regions containing HH~2 (left) and HH~1 (right).
Also, on the H$\beta$ frames we show boxes which include the
brighter regions of HH~1 and HH~2 (boxes B and A, respectively),
which have been used for calculating the H$\alpha$/H$\beta$
ratios shown in Figures 2 to 5. The images are displayed with a logarithmic greyscale.} 
\end{figure}

As discussed in detail by O'Dell et al. (2013), the F656N filter is contaminated by emission from
the [N II] 6548 line, and both the F656N and F487N filters have contributions from the nebular
continuum. Using the fact that at all measured positions within HH 1 and 2, the [N II] 6548/H$\alpha$
ratio does not exceed a value of $\approx 0.35$ (see, e.g., Brugel, B\"ohm \& Mannery 1981a and Solf et al. 1988)
and the transmission curve of the F656N filter (see O'Dell et al. 2013 and the WFC3 Instrument Handbook)
one then finds a peak contribution of $\approx 2$\%\ to the measured flux. For estimating the
effects of the continuum in the F656N and F487N images one can use the continuum and line fluxes
obtained by Brugel, B\"ohm \& Mannery (1981a, b) and the bandpasses of the filters to obtain
estimates of $\approx 0.4$ and 5\%\ (for the F656N and F487N filters, respectively). Therefore,
when interpreting the H$\alpha$/H$\beta$ ratios obtained from our HST images, it is necessary
to keep in mind that there is an uncertainty of $\sim 5$\%\ due to a possible spatial dependence
in the H$\beta$ line to continuum ratio within the F487N filter. As this uncertainty is $\sim 1$ order
of magnitude smaller than the H$\alpha$/H$\beta$ ratio variations described below, we do not
discuss it further.

\section{The H$\alpha$/H$\beta$ ratios}

Figure 2 shows the H$\alpha$ map (right) and H$\alpha$/H$\beta$ ratio map (left) for HH~2.
To avoid having extended regions dominated by noise, in order to calculate the line ratio map it is necessary to place a
lower bound on the line fluxes. We have chosen to calculate the ratios only in regions in which the observed H$\beta$
flux is larger than $I_0=5.4\times 10^{-18}$ erg s$^{-1}$pix$^{-1}$.

\begin{figure*}
\centering
\includegraphics[width=16cm]{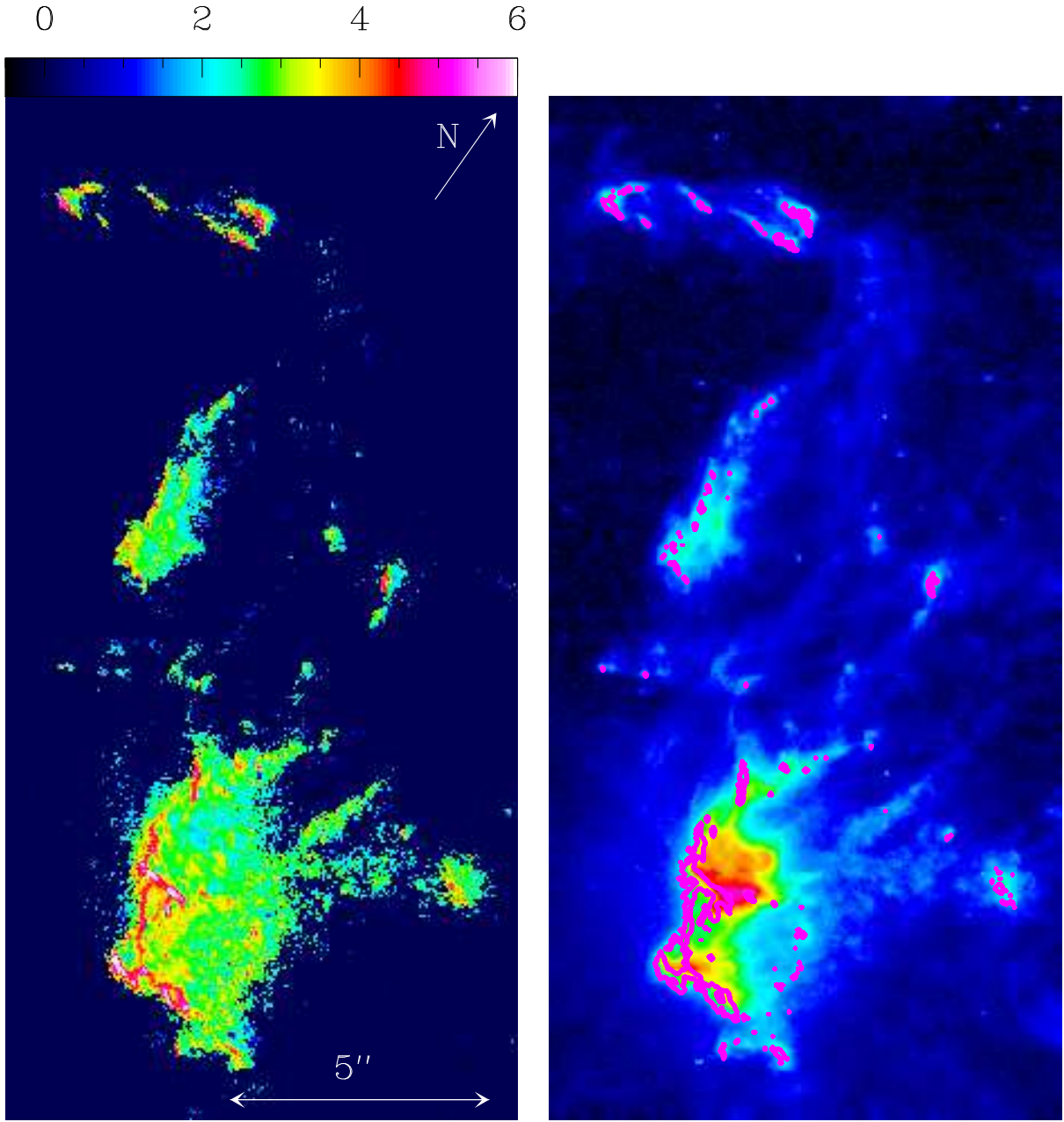}
\caption{H$\alpha$ emission (right, shown with a logarithmic colour scale) and dereddened
H$\alpha$/H$\beta$ ratio (left, shown with the linear colour scale given by the top bar) for
HH~2. The outflow source lies
towards the NNW. On the H$\alpha$ image
(right frame) we have included a (dereddened) H$\alpha$/H$\beta$=4 contour (in purple). This
contour shows that the higher values of the line ratio are distributed in
a ridge along the leading edge of the emitting condensations.} 
\end{figure*}

For calculating the intrinsic H$\alpha$/H$\beta$ ratios we have applied the following reddening correction. We
first calculate the observed ratios for all of the pixels with H$\beta$ intensities larger than $I_0$ (see above)
for the A and B boxes (shown in the bottom frames of Figure 1). For HH 2 we obtain a mean line
ratio $<\rm H\alpha/H\beta>_{obs}=3.82$, and for HH 1 an almost identical
$<\rm H\alpha/H\beta>_{obs}=3.79$ value. Considering an observed line ratio of 3.8 for both objects,
comparing with the case B recombination cascade intrinsic H$\alpha$/H$\beta$ ratio of 2.8 and using
the average Galactic extinction curve, we obtain an $E(B-V)=0.27$ colour excess. This value is somewhat
lower than the $E(B-V)\approx 0.35$ value deduced for HH~2 by Brugel et al. (1981a), using the method
of Miller (1968), based on the fixed ratios between the auroral and transauroral lines of [S II] (i.e., not assuming
a recombination cascade H$\alpha$/H$\beta$ ratio). In order to calculate the dereddened H$\alpha$/H$\beta$
ratios, we therefore multiply the observed ratios by a factor of 2.8/3.8, basically assuming that the extinction
towards HH 1 and 2 is position-independent.

The dereddened H$\alpha$/H$\beta$ ratios of HH~2 (see Figure 2)
have values in the $2\to 6$ range, with the regions of higher values corresponding
to filamentary structures in the leading edge of the emitting knots (i.e., in the edges directed away from the
outflow source). In order to illustrate the positions of
these ``high H$\alpha$/H$\beta$'' regions, we have superimposed an H$\alpha$/H$\beta=4$ contour on the
H$\alpha$ emission map (purple contour in the right frame of Figure 2).

Figure 3 shows the H$\alpha$ map (bottom) and dereddened H$\alpha$/H$\beta$ ratio map (top)
for HH~1.  We have calculated the
ratios only for pixels with an observed H$\beta$ flux larger than $I_0=5.4\times 10^{-18}$ erg s$^{-1}$pix$^{-1}$ (i.e., the
same cutoff used for HH~2, see above). The region with H$\alpha$/H$\beta>4$ is a thin filament
on the E side of the leading edge of HH~1 (see the purple contour on the H$\alpha$ emission map in the bottom
frame of Figure 3). It is clear that HH~1 shows a strong side-to-side asymmetry with respect to the outflow axis, as
the SW region of the leading edge does not show high H$\alpha$/H$\beta$ ratios (see the top frame of Figure 3).
The H$\alpha$ emission also shows a strong side-to-side asymmetry.

\begin{figure}
\centering
\includegraphics[width=9cm]{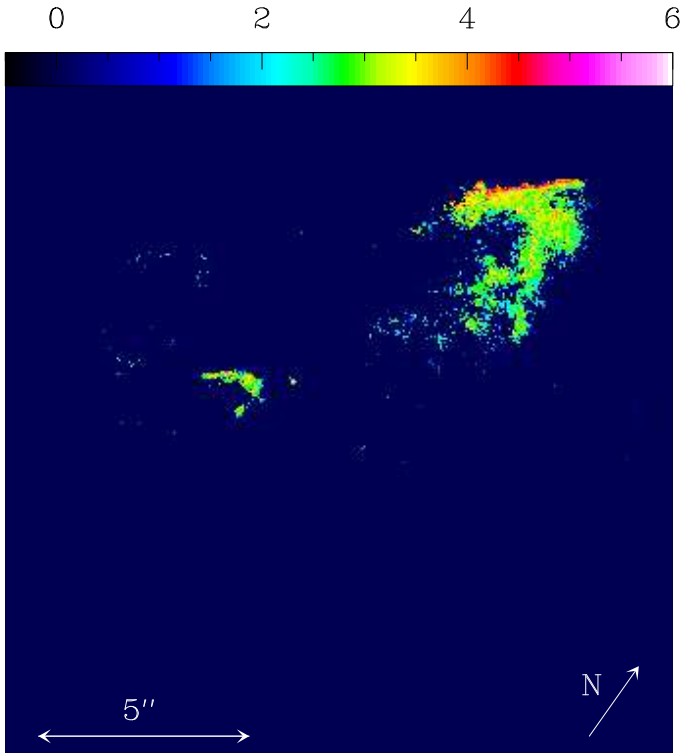}
\caption{H$\alpha$ emission (bottom, shown with a logarithmic colour scale) and dereddened
H$\alpha$/H$\beta$ ratio (top, shown with the linear colour scale given by the top bar) for
HH~1. The outflow source lies
towards the SSE. On the H$\alpha$ image
(bottom frame) we have included a (dereddened) H$\alpha$/H$\beta$=4 contour (in purple). This
contour shows that the higher values of the line ratio are distributed in
a ridge along the E side of the leading edge of HH~1.}
\end{figure}

Figure 4 shows the dereddened H$\alpha$/H$\beta$ line ratio as a function of the (observed) H$\beta$ flux for all of the
pixels with $I_{H\beta}>I_0$ (see above) for HH~1 (top frame) and HH~2 (bottom frame). It is clear that for low
values of the H$\beta$ intensity in both HH~1 and 2 we have a relatively broad distribution of line ratios (the
width of this distribution representing the relatively large errors of the line ratio at low intensities) centered
on the H$\alpha$/H$\beta=2.8$ recombination cascade value. For pixels with brighter H$\beta$ intensities, we
see a distribution of  H$\alpha$/H$\beta$ ratios extending from $\approx 3$ to larger values of
$\sim 5$ (for HH~1) or $\sim 6$ (for HH~2).

\begin{figure}
\centering
\includegraphics[width=8cm]{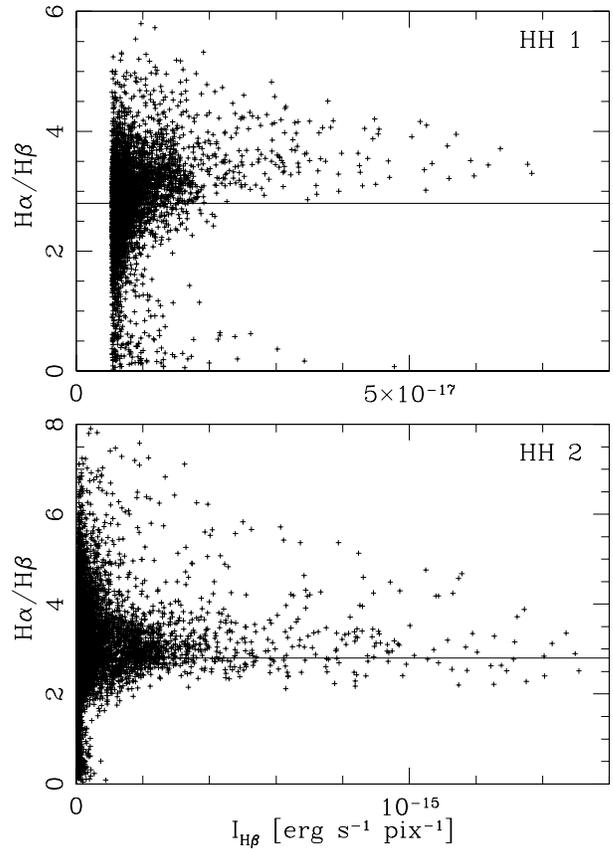}
\caption{Dereddened H$\alpha$/H$\beta$ ratio as a function of the observed H$\beta$
flux for the pixels of HH~1 (top) and HH~2 (bottom). For the lower H$\beta$ intensities we
see that the line ratios straddle a value of $\approx 2.8$ (shown with a horizontal line in both
frames), corresponding to the recombination cascade H$\alpha$/H$\beta$ ratio. For larger
H$\beta$ intensities we find line ratios extending from $\approx 3$ to $\approx 5$ (for HH~1)
or 6 (for HH~2).} 
\end{figure}

This result is seen more clearly in Figure 5, where we show the normalized distributions of the line ratios
of pixels with $I_0<I_{H\beta}<I_1=2.5\times 10^{-17}$ erg s$^{-1}$ pix$^{-1}$ (distribution $f_1$, top frame), of
pixels with $I_1<I_{H\beta}<I_2=4.7\times 10^{-16}$ erg s$^{-1}$ pix$^{-1}$ (distribution $f_2$, center), and of
all pixels with $I_2<I_{H\beta}$ (distribution $f_3$,
bottom frame of Figure 5, with appropriate pixels found only in HH~2).
For both HH~2 (left column) and HH~1 (right column of Figure 5), we see that the distribution $f_1$ of the
lower intensity pixels is approximately symmetrical, centered at an H$\alpha$/H$\beta\approx 2.8$
line ratio. The distributions for higher intensity pixels ($f_2$ and $f_3$, see above and the central and bottom
frames of Figure 5) start at values of H$\alpha$/H$\beta \sim 2$-2.5, have a peak at a line ratio
of $\approx 3.3$ and have a wing extending to H$\alpha$/H$\beta \sim 5$ for HH~1 and $\sim 6$ for
HH~2. In the following section, we show that these high H$\alpha$/H$\beta$ ratios coincide with the
values expected for collisional excitation of the $n=3$ and 4 levels of H.

\begin{figure}
\centering
\includegraphics[width=8cm]{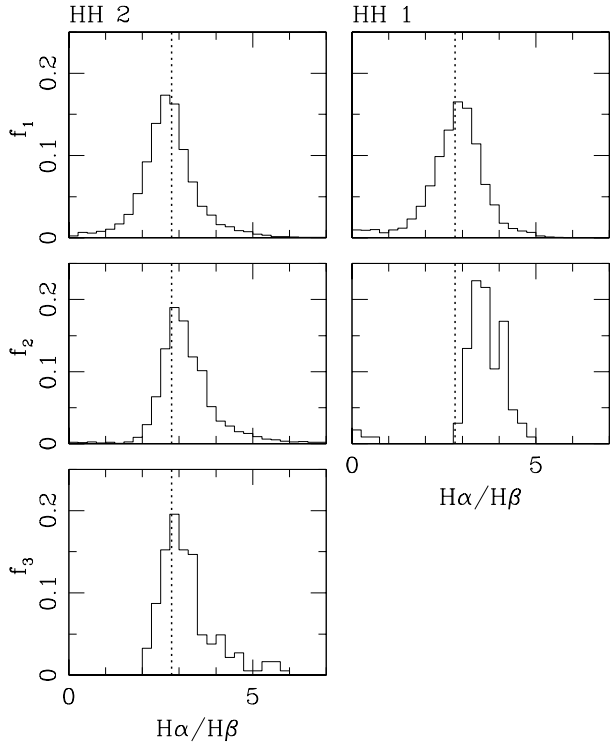}
\caption{Distribution functions for the number of pixels within H$\alpha$/H$\beta$ ratio bins
for HH~2 (left) and HH~1 (right). We show three different distributions: $f_1$ (top) corresponding
to pixels with $I_0=5.4\times 10^{-18}$ erg s$^{-1}$ pix$^{-1}<I_{H\beta}<I_1=2.5\times 10^{-17}$
erg s$^{-1}$ pix$^{-1}$, $f_2$ (center) of pixels with $I_1<I_{H\beta}<I_2=4.7\times 10^{-16}$
erg s$^{-1}$ pix$^{-1}$ and $f_3$ (bottom) of
pixels with $I_2<I_{H\beta}$. The distribution function of the lower intensity pixels ($f_1$,
top frames) is approximately symmetrical, centered at a line ratio of $\approx 2.8$, corresponding
to the recombination cascade value (the dashed, vertical line in all plots corresponds to
H$\alpha$/H$\beta=2.8$). The distribution functions for higher intensity pixels ($f_2$ and $f_3$,
central and bottom frames) all show extended wings to higher values of H$\alpha$/H$\beta$.}
\end{figure}

\section{Discussion}

From our new H$\alpha$ and H$\beta$ HST images we can compute dereddened H$\alpha$/H$\beta$
maps for the brighter regions of HH~1 and 2. For the reddening correction, we assume that
the mean value of the H$\alpha$/H$\beta$ ratio coincides with the recombination cascade
value of 2.8, as found previously by Brugel et al. (1981a), who calculated the reddening correction
with Miller's method, based on the ratios of auroral to transauroral [S~II] lines.

We find that in limited spatial regions the (dereddened) H$\alpha$/H$\beta$ ratio has
values of $\sim 4\to 6$, which are inconsistent with the recombination cascade value. These
high H$\alpha$/H$\beta$ regions are filaments along the leading edges (i.e., the edges away from
the outflow source) of the brighter emitting regions of HH~1 and 2 (see Figures 2 and 3).

Raga et al. (2014) show that the $r_{\alpha\beta}=$H$\alpha$/H$\beta$ ratio for a ``case B''
cascade fed by collisional excitations from the ground state of hydrogen has the approximate form:
\begin{equation}
r_{\alpha\beta}=3.40\,e^{E_{43}/(kT)}+\frac{0.95}{\left(1+{\rm 5\times 10^4\,K}/T\right)^4}\,,
\label{r}
\end{equation}
where $k$ is Boltzmann's constant and $E_{43}$ is the energy difference between the $n=4$
and $n=3$ energy levels (so that $E_{43}/k=7680$~K). The first term of
this functional form has a temperature
dependence derived from the ratio of the $n=1\to 3$ and $n=1\to 4$ collisional excitation
coefficients (assuming temperature-independent collision strengths),
and the second term is a correction necessary to match the results of a 5-level, collisionally
fed cascade matrix description of the hydrogen atom in the $T=10^3\to 10^6$K
temperature range (see Raga et al. 2014). It is clear that the functional
form of $r$ (see equation \ref{r}) has high values for low temperatures, and has an asymptotic
value of 4.35 for $T\to \infty$.

From equation (\ref{r}), one obtains $r(T=1.5\times 10^4 {\rm K})=5.6$ and $r(T=10^5 {\rm K})=3.9$.
Therefore, the wing of the line ratio distributions extending to
H$\alpha$/H$\beta\sim 4\to 6$ (see Figure 5) can straightforwardly be explained as produced in
regions with temperatures in the $1.5\to 10\times 10^4$~K range emitting collisionally excited Balmer lines.

This clear evidence that we are observing collisionally excited Balmer lines together with the fact
that the high H$\alpha$/H$\beta$ regions are restricted to the leading edges of the outward moving
condensations of HH~1 and 2 is quite conclusive evidence that we are observing the region of collisional
excitation of H lines right after the shock waves driven into the surrounding medium by the condensations.
Most of the H$\alpha$ emission, however, comes from a region further away from the shock,
in which the Balmer lines are produced through the standard recombination cascade (as evidenced
by the H$\alpha$/H$\beta\sim 3$ ratios, see Figures 2 and 3).

The theoretical prediction of
these two regions of Balmer line emission (a collisionally excited Balmer line region immediately after
the shock, and the recombination region with Balmer lines dominated by the recombination cascade)
in HH shock wave models is already mentioned by Raymond (1979), and the H$\alpha$
emission from the two regions was studied in more detail by Raga \& Binette (1991). These two regions
are of course present in all shock models (for example, in the plane-parallel, time-dependent
shock models of Te{\c s}ileanu et al. 2009).

In non-radiative shocks observed in some supernovae remnants or in pulsar cometary nebulae, the observed
emission comes exclusively from the region of collisional excitation right behind the shocks (see, e.g.,
the review of Heng 2010). In HH objects, the only previous observational evidence of the emission
from the immediate post-shock region (as opposed to the emission from the recombination region)
were the H$\alpha$ filaments seen in HST images of some bow shocks, notably in the HST images
of HH~47 (Heathcote et al. 1996), HH~111 (Reipurth et al. 1997) and
HH~34 (Reipurth et al. 2002). However, as only H$\alpha$ was observed it was not possible to
guarantee that these filaments did correspond to the region of collisionally excited Balmer lines.

Our new H$\alpha$ and H$\beta$ images for the first time show in a quite conclusive way that
we have a detection of the immediate post-shock region of HH objects (in which H is being collisionally
ionized and the levels of H are being collisionally excited). The detection of this region provides a
clear way forward for developing models of HH bow shocks, in which the position of the shock
wave relative to the recombination region is directly constrained by the observations.

We should note that throughout this paper we have assumed that the exctinction is uniform
over the emission regions of HH~1 and 2. In principle it could be possible that foreground
structures in the vicinity of the objects might produce changes in the extinction on angular
scales comparable to the size of the objects. However, estimates of the density of the
pre-bow shock material of HH~1 and 2 (based on observations of the post-shock density and
on plane-parallel shock models, see, e.g., Hartigan et al. 1987) give values
$\sim 100$-300~cm$^{-3}$. Clearly, such a low density environment will not produce
appreciable extinction on spatial scales comparable to the size of the HH objects. Because
of this, if one wants to attribute the observed changes in the H$\alpha$/H$\beta$ ratio
to an angular dependence of the extinction, it is necessary to assume that still undetected,
sharp-edged, high density regions are present in the immediate vicinity of HH~1 and 2.

\begin{acknowledgements}
Support for this work was provided by NASA through grant HST-GO-13484 from the
Space Telescope Science Institute. AR and ACR acknowledge support from the CONACyT grants
101356, 101975 and 167611 and the DGAPA-UNAM grants IN105312 and IG100214.
\end{acknowledgements}


\end{document}